\documentclass[twocolumn,eqsecnum,nofootinbib]{revtex4}
\usepackage{graphicx}

\def\beq{\begin{equation}}
\def\eeq{\end{equation}}
\def\bea{\begin{eqnarray}}
\def\eea{\end{eqnarray}}

\begin{document}
\title{String reveals secrets of universe}
\author{Yong Seung Cho}
\email{yescho@ewha.ac.kr} \affiliation{Department of Mathematics,
Ewha Womans University, Seoul 120-750, Republic of
Korea}
\author{Soon-Tae Hong}
\email{soonhong@ewha.ac.kr} \affiliation{Department of Science
Education and Research Institute for Basic Sciences, Ewha Womans
University, Seoul 120-750, Republic of Korea}
\date{\today}
\begin{abstract}
Stringy cosmology displays features that are different from
standard cosmology. One may be surprised in that in this scenario
there is no phase transition between the radiation dominated phase
and matter dominated phase and the universe is cyclic similar to
brane cyclic cosmology.
\end{abstract}
\maketitle

Modern mathematical cosmology is based heavily on the
Hawking-Penrose singularity theorem~\cite{hawking70} which states
that once collapse approaches a certain point, evolution to a
singularity is unavoidable. Here the singularity denotes something
similar to a shrinking point beyond which extending the spacetime
is unattainable. The singularity theorem indicates that there was
a singularity at the beginning of our universe which is believed
to have expanded now. In the standard cosmology, the expansion of
the universe from a single point, namely Big Bang is assumed to
have occurred at this singularity.

We now consider phase transition which is assumed to have happened
in the evolution of the universe around 75,000 years after the Big
Bang. Before going into details of the phase transition in
cosmology, we consider the water phase transition between the gas
and liquid phases, which takes place at 100 degrees Celsius as the
temperature of the gaseous water cools down. When either above or
below the critical temperature the water is either gas or liquid,
and at the critical temperature water is in a gas-liquid mixture
form. It is known that in the universe there exist two types of
particles: massive particles such as electrons and massless ones
such as photons, namely quanta of light. In obtaining the
Hawking-Penrose singularity theorem, they have exploited the
so-called strong energy condition.  Assuming that the early
universe was filled with a perfect fluid consisting of massive
particles and/or massless particles and using the strong energy
condition, one could find equations of state for each particle.

Specifically, the equation of state of the massive particle is
different from that of the massless particle, which indicates that
the massive particle phase is not the same as the massless
particle one. In the standard cosmology based on the
Hawking-Penrose singularity theorem and inflationary scenario, it
is believed that, after the Big Bang explosion, a massless
particle (or radiation) dominated phase occurred followed by a
massive particle (or matter) dominated one, even though there was
a hot thermalization period of radiation and matter immediately
after the Big Bang. Moreover, as in the phase transition in water,
a phase transition exists between massive particle and massless
particle phases in the universe.

In the above explanations of the Hawking-Penrose singularity
theorem and the ensuing standard cosmology, they assumed that
massive and massless particles are point-like. Now let us discuss
stringy cosmology which has been developed recently~\cite{cho07}
and is based on the string theory~\cite{pol98}. The string theory
has given us a better understanding of the universe and has
provided an analytical tool for studying the nature of the
universe. It was proposed in the string theory that massive and
massless particles in the universe can be described by vibrating
strings instead of point particles. As in the standard (point
particle) cosmology, using the strong energy condition modified by
stringy corrections, one can find the stringy singularity theorem,
which is similar to the Hawking-Penrose singularity theorem except
for the fact that evolution to a singularity occurs at different
spacetime~\cite{cho07}.

Exploiting the stringy strong energy condition together with the
assumption that the early universe was occupied with a perfect
fluid of massive stringy particles and/or massless stringy
particles, one obtains equations of state for each stringy
particle. Surprisingly, the massive stringy particles and massless
stringy particles produce the same equation of state. This implies
that both massive stringy particles and massless stringy particles
are in the same phase and thus there is no phase transition in the
stringy universe~\cite{cho07}. Here, we emphasize that the
equations of state obtained from the standard and stringy
cosmologies govern the evolution of the universe, since these
equations originate from the initial conditions of the (stringy)
singularity theorems. Immediately after the Big Bang explosion, in
the stringy cosmology, the massless stringy particle dominated
phase and the massive stringy particle dominated phase took place
simultaneously. In the stringy universe, a massless-massive
stringy particle mixture state without any phase transitions
exists. These features in the stringy universe
are drastically different from those in the standard point
particle cosmology.  In fact, these differences originate from the
lone fact that the particles are composed of strings instead of
points.

In the standard cosmology, after the radiation and matter
dominated phases a dark energy dominated phase exists. In this
inflationary scenario, a sequence of epoches occurred: Big Bang,
radiation dominated phase, matter dominated phase, and dark energy
dominated phase. In brane cyclic cosmology, which is also based on
the string (or D brane) theory, the universe is assumed to be
cyclic in evolution: Big Bang, radiation dominated phase, matter
dominated phase, dark energy dominated phase, Big Crunch, and
again Big Bang~\cite{steinhardt}. In the stringy cosmology,
however, the dark energy dominated phase still remains an open
problem and could be properly adopted. Moreover, both in the
standard and stringy cosmologies exploiting the (stringy)
singularity theorems, there exists an initial state with a
negative expansion rate and the ensuing evolution to the
singularity. This phenomena can be interpreted as the Big Crunch,
even though in the standard cosmology the Big Crunch is excluded.
One can thus claim that the stringy cosmology is also cyclic,
similar to the brane cyclic cosmology, but modified: Big Bang,
radiation-matter mixture phase, dark energy dominated phase, Big
Crunch, and again Big Bang, which is consistent with the fact that
both the stringy and brane cyclic cosmologies have the same
characteristics in that these scenarios use the string features.

It is known that protons and neutrons consist of particle
constituents: massless gluons and massive quarks. Recently, the
Large Hadron Collider (LHC) has been built at the European
Organization for Nuclear Research (CERN) near Geneva, Switzerland.
In particular, the Alice detector of the LHC is scheduled to
detect the so-called quark-gluon plasma state, which is assumed to
exist in an extremely hot soup of quarks and gluons. Both in the
standard and stringy cosmologies, this quark-gluon plasma state is
supposed to occur immediately after the Big Bang of the tiny early
universe manufactured in the LHC. However, there are drastically
different ensuing processes in these two scenarios. Namely, in the
standard cosmology, the quark-gluon plasma state can exist shortly
and disappear eventually to enter the radiation dominated phase,
while in the stringy cosmology the quark-gluon plasma state can
develop into particles such as protons and neutrons and sustain
the radiation and matter mixture phase. We recall that as far as
radiation and matter are concerned, the mixture of these two
coexist in the present universe. It is expected that the Alice
will be able to detect the procedure of particle states along with
the evolution of the tiny universe planned to occur at the LHC and
it will be able to determine which cosmology is viable.

\end{document}